\documentclass[fleqn,aps,prb,preprint,a4paper,onecolumn]{revtex4}
\pdfoutput=1
\usepackage[]{amsmath}
\usepackage{amssymb}
\usepackage[dvips]{graphicx}
\usepackage{color}
\usepackage{tabularx}
\usepackage{mathtools}
\usepackage{algpseudocode}
\usepackage{enumitem}
\usepackage{multirow}
\usepackage{epstopdf}  
\usepackage{bm}
\usepackage{float}
\usepackage{threeparttable}


\begin{document}

\title{A Generalizable Machine-learning Potential of Ag-Au Nanoalloys and its Application on Surface Reconstruction, Segregation and Diffusion}
\author{Yinan Wang}
\affiliation{Songshan Lake Materials Laboratory, Dongguan, Guangdong 523808, People's Republic of China\\
Institute of Physics, Chinese Academy of Sciences, Beijing 100190, People's Republic of China}
\author{LinFeng Zhang}
\affiliation{Beijing Institute of Big Data Research, Beijing 100871, China}
\author{Ben Xu}
\affiliation{Graduate School, China Academy of Engineering Physics, Building 9, Beijing 100193, People's Republic of China}
\author{Xiaoyang Wang}
\email{xiaoyanglanl@gmail.com}
\affiliation{Laboratory of Computational Physics, Institute of Applied Physics and Computational Mathematics, Huayuan Road 6, Beijing 100088, People's Republic of China}
\author{Han Wang}
\email{wang_han@iapcm.ac.cn}
\affiliation{Laboratory of Computational Physics, Institute of Applied Physics and Computational Mathematics, Huayuan Road 6, Beijing 100088, People's Republic of China}

\begin{abstract}
Owing to the excellent catalysis properties of Ag-Au binary nanoalloy, nanostructured Ag-Au, such as Ag-Au nanoparticles and nanopillars, have been under intense investigation. 
To achieve high accuracy in molecular simulations of the Ag-Au nanoalloys, the surface properties are required to be modeled with first-principles precision. In this work, we propose a generalizable machine-learning interatomic potential for the Ag-Au nanoalloys based on deep neural networks, trained from a database constructed with the first-principle calculations. 
This potential is highlighted by the accurate prediction of Au (111) surface reconstruction and the segregation of Au towards the Ag-Au nanoalloy surface, where the empirical force field failed in both cases. Moreover, regarding the adsorption and diffusion of adatoms on surfaces, the overall performance of our potential is better than the empirical force fields. 
We stress that the reported surface properties are blind to the potential modeling in the sense that none of the surface configurations is explicitly included in the training database, therefore, the reported potential is expected to have a strong ability of generalization to a wide range of properties and to play a key role in the investigation of nanostructured Ag-Au evolution, where the accurate descriptions of free surfaces are necessary.
\end{abstract}

\maketitle

\setlength{\parskip}{.5em}

\section{Introduction}
Noble metals like Ag, Au, and Ag-Au alloys are widely used in energy utilization and catalysis science.
Au is a great catalyst for oxidation reactions \cite{bond2006catalysis}, especially in the oxidation of carbon monoxide (CO) \cite{haruta1987novel,haruta2005gold} and the selective oxidation of alcohols \cite{funakawa2004selectivity,zope2010reactivity,xu2010vapour}. 
In electrochemistry, the Au (111) surface is an active electrode material for the oxidation of CO \cite{rodriguez2009unusual,rodriguez2010carbon}. 
In addition to the Au (111) surface, the nanoporous gold (npAu) has also attracted attentions because of its high reactivity and high catalytic activity \cite{zielasek2006gold,xu2007low,wittstock2012nanoporous,wittstock2014catalysis}. 
The residual silver remaining in npAu after a dealloying process will enhance the catalytic activity of npAu, which has been demonstrated by several works \cite{wang2013catalytic,moskaleva2011silver,krekeler2017silver}. 
Regarding the Ag-Au binary systems, the superior property of Ag-Au nano-particles has attracted extensive attentions in fields such as optics, catalysis, and electronics~\cite{aiken1999review,murphy2005anisotropic}. 
Besides, the Ag-Au nanoparticles share the outstanding plasmonic response of Ag~\cite{yang2014galvanic} and the chemical stability of Au~\cite{krishnan2018seed}. 

Molecular dynamics is
a powerful tool for revealing the microscopic picture of Ag-Au-related phenomena, such as, e.g., the reconstruction of the Au (111) surface~\cite{ravelo1989molecular,ercolessi1987surface}, the segregation of the Ag-Au alloy on the surface~\cite{bozzolo2007atomistic,deng2011ag}, as well as the diffusion of adatoms on the surfaces~\cite{bon2019reliability,ahmad2019template}, etc..
In an MD simulation, an atomic system is typically evolved with the interatomic forces generated on-the-fly using either first-principles electronic structure methods such as the density functional theory (DFT)~\cite{PhysRev.136.B864,kohn1965self}, or empirical force fields (EFFs).
DFT-based simulation is more accurate and reliable, but the accessible temporal and spacial scales are limited by the prohibitive computational cost. 
On the other hand, EFFs are much more efficient, and hence are used more frequently in large-scale simulations, but their accuracy is often in question, particularly when used for describing multi-component alloy systems, or complex defects. 

Recent development of machine learning (ML) techniques\cite{thompson2015spectral,shapeev2016moment,behler2007generalized,bartok2010gaussian,chmiela2017machine,schutt2017schnet,smith2017ani,han2017deep,zhang2018deep,zhang2018end} gives a feasible solution to the issue of the accuracy versus efficiency dilemma between DFT and EFFs.
Upon training, ML-based interatomic potential models hold the promise of having the efficiency comparable with EFFs without loss of DFT accuracy.
In this work, a smooth edition of Deep Potential (DP)\cite{zhang2018end} is employed to generate the interatomic potential energy model for the Ag-Au alloys in the full concentration space.
To generate an optimal set of training data, we use a concurrent learning strategy named Deep Potential Generator (DP-GEN) \cite{zhang2019active}, which has already had successful applications in the Al-Mg alloys\cite{zhang2019active}, the Al-Mg-Cu alloys\cite{Jiang2020AccurateDP}and high-pressure water phases\cite{zhang2021phase}, etc..
In detail, the dataset at the level of DFT with PBE exchange-correlation (xc) functional (DFT-PBE)~\cite{Perdew1996PBE} for the Ag-Au alloys was constructed with the DP-GEN scheme, with consideration of the entire concentration space and of the temperature range $50\leq T\leq 2579$~K and pressure range $0 \leq P \leq 5$~GPa.

Although the PBE exchange-correlation functional is frequently employed in metals and alloys, its predicted lattice parameter and surface formation energy (see Sec.~\ref{sec:results} for more details) of the Ag-Au alloy system made deviate from experimental values, thus the DP model trained with DFT-PBE database (denoted as DP-PBE) is not expected to be reliable. 
One possible way of fixing the issue is to introduce an additional D3 dispersion correction\cite{grimme2010consistent}, and the corrected properties are found to be in good agreement with the experimental records~\cite{reckien2012implementation,suh1988high}. 
Besides, DFT-PBE-D3 gives qualitatively reasonable results for the surface adsorption behaviors \cite{reckien2012implementation}, and is widely used in investigations of the Ag segregation at the Ag-Au surface \cite{hoppe2017first} and the adatom adsorption and diffusion on the Ag or Au surfaces \cite{bon2019reliability}. Therefore, a DP model honest to DFT-PBE-D3 is required for the prediction of surface behavior in larger scales. 

In order to generate such a model, instead of recalculating the entire datebase with PBE-D3 all over again, we calculate only a small portion of the database (around 0.6\%) and use a transfer-learning scheme to obtain a new model in agreement with the precision with PBE-D3 correction. The transfer learning scheme possesses a similar spirit as the one used to transfer knowledge in the ANI models~\cite{smith2019approaching}, so that a significant computational cost is saved. And the resultant model denoted as DP-PBE-D3.

The DP-PBE and DP-PBE-D3 models show great consistency with DFT-PBE and DFT-PBE-D3 datasets, respectively.
The basic properties (including formation energy, volume for equilibrium state, equation of state, elastic moduli, vacancy and self-interstitial formation energy, and unrelaxed surface formation energy) of the structures obtained from the Materials Project (MP) database predicted by DP-PBE and DP-PBE-D3 models are comparable with DFT-D3 and DFT-PBE-D3 results, respectively. 
It is noted that these structures are not explicitly presented in the training database during the DP-GEN procedure.

Moreover, since the DP-PBE-D3 model is trained with D3 correction, it is able to successfully predicted the Au (111) surface reconstruction and the Au enrichment on the (111) surface of Ag-Au alloys that the EFF failed to do, and the DP-PBE-D3 model predicted more accurate adsorption energies and diffusion barriers than the EFFs. 

The paper is organized as following: in Sec.~\ref{sec:method}~, a brief introduction of the DP-GEN scheme and the transfer learning method of the Ag-Au system is provided. In Sec.~\ref{sec:results}~, the basic properties predicted by DFT-PBE, DFT-PBE-D3, DP-PBE, DP-PBE-D3 and an EFF are compared. 

Having proved the reliability of these models, we address three specific problems key to the Ag-Au surface properties: the Au (111) reconstruction, the Au enrichment on Ag-Au alloy (111) surface and the adsorption and diffusion of the adatoms on surfaces.

\section{Method}
\label{sec:method}
\subsection{\label{sec:DPGEN}DP-GEN scheme}
DP-GEN is a concurrent learning scheme that generates the training dataset for DP in an iterative way~\cite{zhang2019active,zhang2020dpgen}.
Starting with an ensemble of initial guesses of the DP models, each DP-GEN iteration contains three steps, \textit{exploration, labeling, and training}.
In the \textit{exploration} step, the configuration space and chemical space are explored by a DP-based sampler, which, in the current work, is the DP-based MD (DPMD) conducted at the relevant thermodynamics states. 
The maximal standard deviation among the forces predicted by four DP models serves as an error indicator. 
Then the configurations along the DPMD trajectories are categorized  according upper and low error bound. Those configurations below the lower bound and above the upper bound are omitted in the \textit{labeling} step. While those configurations between the bounds are selected as candidates of the database.

In the \textit{labeling} step, the energy, the virial tensor, and the atomic forces of the candidate configurations are calculated by DFT-PBE. 
The labelled configurations are added to the training data set, and in the next iteration, four new DP models are trained with the updated data set. 
The DP-GEN iterations converge when the full sample space is explored and most of the configurations are classified as accurate.
Detailed information of the DP-GEN scheme are found in Refs.\cite{zhang2019active,zhang2020dpgen}.

The DP-GEN scheme is implemented by the open-source package \texttt{DP-GEN}\cite{zhang2020dpgen}.
Details on initial data set, exploration, labeling, and training steps are provided below.

\paragraph{Initial data set.}

The initial data set is used to train the ensemble of initial guesses of the DP models. 
In this work, for pure Ag and Au, as well as Ag-Au binary alloys, the initial data sets are prepared in three steps. 
First, for each system, equilibrium  body-centered-cubic (BCC), hexagonal-close-packed (HCP), and face-centered-cubic (FCC) structures, within $2\times 2\times 2$ super-cells, are  constructed. 
For the Ag-Au alloy, all possible concentrations are considered, and, subject to each concentration, the lattice sites are randomly occupied.
Second, the atomic coordinates are perturbed by a random number in the range $\pm$0.01~\AA, and the cell vectors are randomly perturbed in the range of $\pm$~3\% of the original matrix.
Besides, the cells are compressed to cover the high-pressure configurations, with the compression ratio ranging from 0.84 to 0.98 of the equilibrium volume, with a step of 0.02 for pure Ag and Au, and a step of 0.04 for Ag-Au alloys.
Finally, short trajectory {\it ab initio} molecular dynamics (AIMD) simulations with 5 steps are conducted starting from the perturbed configurations, and the the energy, force, and virial tensor along the trajectories are collected as the initial data sets. 

\paragraph{Exploration.}

The \texttt{LAMMPS} package~\cite{plimpton1995lammps} compiled with the \texttt{DeePMD-kit}~\cite{wang2018deepmd} support is employed to perform DP based MD simulations for the exploration of the configuration and content space. 
We consider both bulk and surface structures to correctly describe the basic properties and the surface related properties, such as the surface structures, surface segregation, as well as the adsorption and diffusion of adatom on surfaces.

The perturbed and compressed BCC, HCP, and FCC structures serve as the initial configurations of the DPMD simulations in the first 32 DP-GEN iterations. 
Isothermo-isobaric (NPT) MD simulations are conducted at temperatures ranging from 50.0 to 2346.367~K for Ag, and to 2579.763~K for Au, Ag-Au alloy, respectively. 
The pressures are sampled ranging from 0.0001 to 5.0~GPa.

The surface configurations are explored by DPMD in the next 16 iterations. 
The initial configurations are created with low Miller indexes of FCC and HCP structures. 
We consider \{100, 110, 111\} and \{001, 100, 110\} for FCC and HCP, respectively. 
The surface structures are constructed so that the thickness is at least 10~\AA, and the surface normal vectors are placed along the $z$ direction of the simulation cell. 
The generated FCC surface structures are further replicated along the $x$ direction by twice.
The HCP surface structures are copied along both $x$ and $y$ directions by twice.
Perturbations similar to the bulk structures (atomic coordinates and cell vectors) are applied to the surface structures. 
The cells are scaled with scaling factors ranging from 0.96 to 1.06, with a step of 0.02. 
The temperature range for canonical ($NVT$) MD simulations is the same  as that of the bulk structures. 
 
We set the lower and upper error bounds to 0.05 eV/\AA\ and 0.20 eV/\AA, respectively.
The candidate configurations are picked if their model deviation of force lies between the lower and upper bounds. 
These configurations will be labeled and added to enlarge the data set. 
During the whole DP-GEN process, about 49.67 million Ag-Au alloy configurations are explored, while only a small portion of them (i.e.~44904, $\sim0.0904\%$) are selected for labeling.  

\paragraph{Labeling.}
The labels of the candidate configurations, i.e.~the energy, force, and virial tensor, are computed by DFT.
The DFT calculations were conducted using the Vienna {\it ab initio} simulation package (\texttt{VASP})\cite{kresse1996efficient,kresse1996efficiency}, with the generalized gradient approximation (GGA) and the Perdew-Burke-Ernzerhof (PBE)\cite{Perdew1996PBE}
exchange-correlation functional. 
The projector-augmented-wave (PAW)
method\cite{blochl1994projector,PhysRevB.59.1758} is used and the energy cut-off of the planewave basis set is set to 650~eV. 
The  Brillouin zone is sampled by the Monkhorst-Pack method\cite{monkhorst1976special} with a grid spacing of  0.1\AA$^{-1}$.
The convergence criterion for the energy is $1 \times 10^{-6}$ eV.

\paragraph{Training.}
The DP-PBE model is constructed using a smooth edition of the Deep Potential model\cite{zhang2018end}.
The \texttt{DeePMD-kit} package\cite{wang2018deepmd} is used for training.
In each iteration, four models are trained simultaneously using the same data set, with the only difference being the random seeds employed to initialize the model parameters. 
The sizes of the embedding and fitting nets are set to $(25, 50, 100)$ and $(240, 240, 240)$, respectively.
The cut-off radius is set to 6~\AA. 
The Adam stochastic gradient descent method\cite{Kingma2015adam} with the default hyper-parameter settings provided by the \texttt{TensorFlow} package\cite{abadi2015tensorflow} was used to train the DP models.
The starting and final learning rate was $ 1\times10^{-3} $ and $ 5\times10^{-8} $, respectively. 
In each DP-GEN iteration the DP model is trained with $ 4.0\times10^{5 
}$ steps. 
After the DP-GEN iterations converge, the final productive models are trained with $ 8.0\times10^{6}$ steps.

\subsection{\label{sec:translearn}Transfer learning}

To get better descriptions of the lattice parameter and surface formation energy, and for further researches in the surface reconstruction, surface segregation and the adatom adsorption and diffusion on surfaces, a DP model based on PBE-D3 accuracy is necessary. 
Instead of conducting the DP-GEN scheme from scratch for DFT-PBE-D3, we propose the following transfer learning procedure to generate a DP model with DFT-PBE-D3 accuracy based on the knowledge encoded in the DP-PBE model.

The transfer learning procedure is initialized by randomly picking a small amount (100 in this work) data from the DFT-PBE labeled dataset, relabeling with D3 correction and training a set of DP-PBE-D3 models from the dataset. 
The parameters of DP-PBE-D3 model are initialized by the DP-PBE model. 
During training, the parameters of the whole embedding net and the hidden layers of the fitting net are fixed, and only the parameters of the output layer of the fitting net are trainable. 
We have explored the possibility of letting the last (out-most) hidden layer of the fitting net to be trainable and found almost no difference with the case of fixed hidden layers.
In what follows a DFT-PBE-D3 labeled dataset is generated by an iterative scheme, and just like the DP-GEN scheme each iteration is composed by exploration, labeling, and training steps.
In the exploration step, the model deviation of DP-PBE-D3 models on the DFT-PBE labeled dataset (labels are ignored) is calculated. 
The data with model deviations smaller than 0.05~eV/\AA\ are excluded, and then at most 100 data are randomly picked from the remained  dataset (those data with model deviation larger than 0.05~eV/\AA), and are sent to the labeling step.
In the labeling step, the D3 dispersion correction is added to the labels, and the labeled data are added to the DFT-PBE-D3 labeled dataset.
Finally, in the training step, the same training strategy as that used in the initialization stage of the transfer learning is adopted.

The transfer learning method is efficient, because one does not need to conduct the DP-GEN from scratch or calculated the D3 correction for all DFT-PBE labeled data, which substantially saves the computational cost on DFT calculations.
For pure Ag, pure Au, and Ag-Au alloy, merely 144, 155 and 275 structures were selected to be labeled with D3 correction, respectively.
The accuracy of the transferred DP-PBE-D3 model will be shown to be in good agreement with DFT-PBE-D3, and will be demonstrated in detail in Sec.~\ref{sec:results}.

\begin{center}
\begin{table}
\setlength{\abovecaptionskip}{0.cm}
\setlength{\belowcaptionskip}{0.9cm}
\caption{Equilibrium properties of pure Ag and Au: lattice parameter $a$, energy per atom in bulk $E_{0}$, independent elastic constants $C_{11}$, $C_{12}$, and $C_{44}$, Bulk modulus $B_v$, shear modulus $G_v$, Young's modulus $E_v$, possion's ratio $\nu$, vacancy formation energy $E_{vf}$, and interstitial formation energies $E_{if}$ for octahedral interstitial (oh) and tetrahedral interstitial (th). Both FCC and HCP structures are considered. }\label{tab:pureAgAu}
\resizebox{\textwidth}{!}{
\begin{threeparttable}
\begin{tabular}{c|cccccc|ccccc}
\hline
\hline
\multirow{2}{*}{Ag} & \multicolumn{6}{c|}{FCC} & \multicolumn{5}{c}{HCP} \\ 
\cline{2-12}																							
	 &	EAM-Zhou	&	DFT-PBE	&	DP-PBE	&	DFT-PBE-D3	&	DP-PBE-D3	&	Expt. 	&	EAM-Zhou	&	DFT-PBE	&	DP-PBE	&	DFT-PBE-D3	&	DP-PBE-D3	\\
\hline																							
$a$~(\AA)	&	4.09	&	4.15	&	4.15	&	4.07	&	4.07	&	4.09$^{a}$	&	2.88	&	2.93	&	2.93	&	2.88	&	2.87	\\
$E_{0}$ (eV/atom)	&	-2.85	&	-2.719	&	-2.72	&	-3.135	&	-3.135	&	-2.850$^{b}$	&	-2.849	&	-2.715	&	-2.716	&	-3.202	&	-3.203	\\
$C_{11}$ (GPa)	&	124.95	&	117.77	&	113.04	&	137.64	&	129.29	&	124.00$^{c}$	&	153.05	&	123.96	&	138.7	&	145.82	&	160.59	\\
$C_{12}$ (GPa)	&	93.93	&	78.29	&	82.26	&	90.76	&	92.99	&	93.40$^{c}$	&	92.82	&	85.5	&	80.16	&	94.75	&	89.06	\\
$C_{44}$ (GPa)	&	49.71	&	44.26	&	46.75	&	52.65	&	53.53	&	46.10$^{c}$ 	&	18.35	&	18.63	&	17.76	&	21.68	&	17.67	\\
$B_v$ (GPa)	&	104.27	&	91.46	&	92.52	&	106.38	&	105.09	&	104.25$^{c}$	&	105.07	&	90.62	&	94.13	&	104.55	&	101.38	\\
$G_v$ (GPa)	&	36.03	&	34.45	&	34.21	&	40.96	&	39.38	&		&	30.7	&	25.26	&	27.87	&	30.82	&	31.32	\\
$E_v$ (GPa)	&	96.93	&	91.83	&	91.36	&	108.91	&	105.02	&		&	83.93	&	69.33	&	76.1	&	84.19	&	85.2	\\
$\nu$	&	0.35	&	0.33	&	0.34	&	0.33	&	0.33	&		&	0.37	&	0.37	&	0.37	&	0.37	&	0.36	\\
$E_{vf}$ (eV) 	&	1.1	&	0.88	&	0.74	&	1.23	&	1.01	&	1.10$^{f}$ 	&	1.16	&	0.74	&	0.74	&	1.13	&	1.01	\\
$E_{if}$ (oh) (eV) 	&	2.64	&	2.88	&	2.67	&	3.23	&	2.98	&		&	1.81	&	2.72	&	2.7	&	3.12	&	2.9	\\
$E_{if}$ (th) (eV) 	&	3.67	&	3.31	&	3.11	&	3.78	&	3.49	&		&		&		&		&		&		\\
\hline																							
\hline																							
\multirow{2}{*}{Au} & \multicolumn{6}{c|}{FCC} & \multicolumn{5}{c}{HCP} \\ 																							
\cline{2-12}																							
	 &	EAM-Zhou	&	DFT-PBE	&	DP-PBE	&	DFT-PBE-D3	&	DP-PBE-D3	&	Expt. 	&	EAM-Zhou	&	DFT-PBE	&	DP-PBE	&	DFT-PBE-D3	&	DP-PBE-D3	\\
\hline																							
$a$~(\AA)	&	4.08	&	4.16	&	4.16	&	4.10	&	4.10	&	4.07$^{d}$	&	2.88	&	2.91	&	2.95	&	2.87	&	2.89	\\
$E_{0}$~(eV/atom)	&	-3.93	&	-3.217	&	-3.216	&	-3.877	&	-3.874	&	-3.930$^{e}$	&	-3.929	&	-3.211	&	-3.213	&	-3.871	&	-3.871	\\
$C_{11}$~(GPa)	&	186.78	&	159.67	&	143.23	&	166.28	&	152.33	&	183.00$^{c}$	&	212.29	&	199.03	&	159.1	&	207.02	&	170.39	\\
$C_{12}$~(GPa)	&	157.68	&	130.13	&	119.71	&	144.75	&	128.03	&	159.00$^{c}$	&	155.76	&	126.46	&	118.21	&	147.92	&	125.43	\\
$C_{44}$~(GPa)	&	45.92	&	33.52	&	33.87	&	36.37	&	35.81	&	45.40$^{c}$	&	17.88	&	19.68	&	18.9	&	8.61	&	8.13	\\
$B_v$~(GPa)	&	167.38	&	139.98	&	127.55	&	151.92	&	136.13	&	180.3$^{c}$	&	166.12	&	138.9	&	133.69	&	149.96	&	132.15	\\
$G_v$~(GPa)	&	33.37	&	26.02	&	25.03	&	26.13	&	26.35	&		&	28.66	&	27.65	&	22.43	&	23.67	&	19.17	\\
$E_v$~(GPa)	&	93.87	&	73.51	&	70.48	&	74.13	&	74.25	&		&	81.32	&	77.78	&	63.74	&	67.46	&	54.87	\\
$\nu$	&	0.41	&	0.41	&	0.41	&	0.42	&	0.41	&		&	0.42	&	0.41	&	0.42	&	0.43	&	0.43	\\
$E_{vf}$~(eV) 	&	0.98	&	0.36	&	0.56	&	0.75	&	1.1	&	0.90$^{f}$	&	1.06	&	0.64	&	0.57	&	0.8	&	0.84	\\
$E_{if}$~(oh)~(eV) 	&	2.5	&	2.78	&	2.42	&	2.84	&	2.47	&		&	2.03	&	2.76	&	2.64	&	2.74	&	2.29	\\
\hline																							

\end{tabular}
 \begin{tablenotes}
        \footnotesize
        \item[a] Reference\cite {Compton2014C}.
        \item[b] Reference\cite {smithells2013metals}.
        \item[c] Reference\cite {simmons1965single}.
        \item[d] Reference\cite {pearson1958lattice}.
        \item[e] Reference\cite {johnson1988analytic}.
        \item[f] Reference\cite {siegel1978vacancy}.
      \end{tablenotes}
\end{threeparttable}}
\end{table}
\end{center}

\section{Results and discussion}\label{sec:results}

In this part, we will firstly compare the reliability of the DP-PBE model and the DP-PBE-D3 model in predicting  basic properties. 
Then we will apply the DP-PBE-D3 model to three representative problems: the Au (111) surface reconstruction, surface segregation, and the adsorption energy and diffusion barriers of adatoms on the surfaces. 

\subsection{\label{sec:basic}Basic properties}

In this section, the basic properties are investigated, including the formation energy, equilibrium volume, equation of state (EOS), elastic modulus, vacancy formation energy, as well as interstitial formation energies. 
For pure Ag and Au, properties of FCC and HCP structures are compared. For the Ag-Au alloys, 7 crystalline structures from the Materials Project (MP)\cite{jain2013commentary} database are selected (MP-1183137, MP-1183205, MP-1183214, MP-1183224, MP-1183227, MP-1229092, MP-985287), corresponding to relative Ag concentrations ranging from $\sim25\%$ to $\sim75\%$. 
An EFF developed by Zhou et.al.~\cite{Zhou2004Misfit} using the embedded atom method (EAM) is denoted by EAM-Zhou and used for comparison.
An automatic workflow implemented in \texttt{DP-GEN}\cite{zhang2020dpgen} is employed to facilitate the computations and analyses of the the properties. 
The DFT calculations are performed with the \texttt{VASP} package\cite{kresse1996efficient,kresse1996efficiency}, and the EAM calculations are performed with \texttt{LAMMPS}\cite{plimpton1995lammps}.
To generate the vacancy and interstitial structures, all non-equivalent defect structures are scanned by the the python package \texttt{pymatgen}\cite{ong2012python} and the interstitial finding tool (\texttt{InFit})\cite {10.3389/fmats.2017.00034}.   
All the structures are visualized by the Open Visualization Tool (ovito) package\cite{stukowski2009visualization}.

\paragraph{Pure Ag and Au}

The equilibrium properties of pure Ag and Au with FCC and HCP structures are listed in Table~\ref{tab:pureAgAu}, including the lattice parameter, energy per atom in bulk, elastic constants, vacancy formation energy and interstitial formation energy. 
It is clearly shown that the PBE-D3 correction resulted in more accurate lattice parameter, energy per atom in bulk and elastic constants compared to the experimental results. 
The defect formation energy is calculated by 
\begin{eqnarray}
E_{df} = E_{d}(N_{d})-N_{d}E_{0}, 
\end{eqnarray}
where $d=v$ or $i$, representing vacancy or interstitial defects. 
$E_{d}$ is the relaxed energy of a defective structure with $N_{d}$ atoms, while $E_{0}$ denotes the energy per atom in perfect lattice. 
Comparing with the DFT-PBE results, the DFT-PBE-D3 described equilibrium properties are closer to the experimental results, indicating the necessity of the D3 correction. For all the properties in Table~\ref{tab:pureAgAu}, the DP-PBE model predicted results in good agreement with the DFT-PBE results, and the results of the DP-PBE-D3 model are consistent with the DFT-PBE-D3 results. 
The EAM-Zhou predictions agree quite well with the experimental results, because that it was fitted explicitly against the experimental measurements of these properties.

\begin{figure}
\begin{center}
\includegraphics[width=0.98\textwidth]{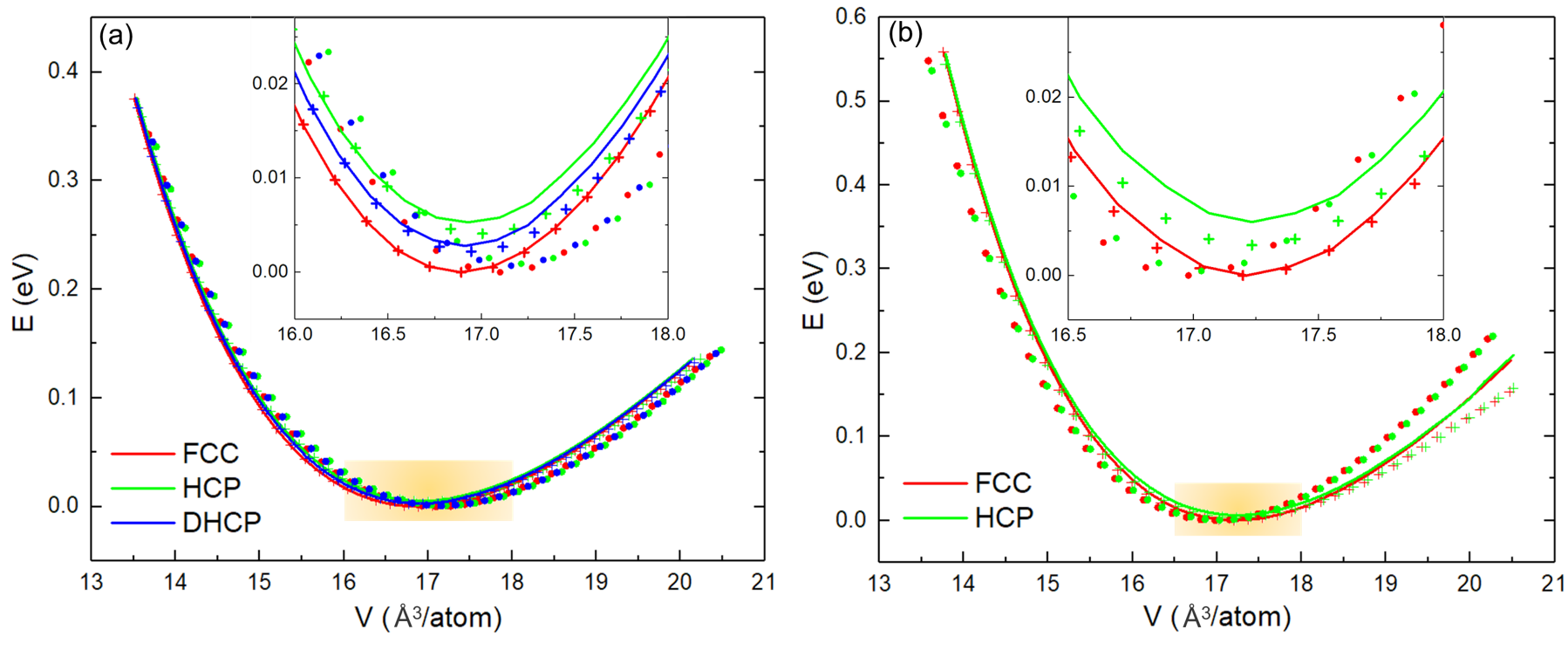}\\[5pt]  
\caption{EOS of (a) Ag and (b) Au. Solid lines, cross points, and dots denote DFT-PBE-D3, DP-PBE-D3, and EAM-Zhou results, respectively. 
Yellow bars denote the volume ranges close to the equilibrium volume and the details around local mimima are amplified in the inset plots.
}\label{fig:eos}
\end{center}
\end{figure}

Fig.~\ref{fig:eos} presents the DFT-PBE-D3, DP-PBE-D3, and EAM-Zhou descriptions for the EOS of Ag (a) and Au (b). 
All energies presented in the figure are relative to the energy of the most stable state, i.e. the equilibrium energy of the FCC phase.
The DP-PBE-D3 model reproduces well the DFT-PBE-D3 EOS for the structures of FCC and HCP for both Ag and Au, together with double-hexagonal-closed-packed (DHCP) for Ag.
As shown in the amplified local details in Fig.~\ref{fig:eos}, the energy minima of DHCP and HCP in Ag are 2 meV/atom and 4 meV/atom higher than the FCC, and DP-PBE-D3 can successfully reproduce this critical relative stability.

Similarly, the EAM-Zhou potential also gives a good description of the relative stability, predicting the FCC phase as the most stable structure. 
However, the energy minima of DHCP and HCP in Ag predicted by EAM-Zhou are respectively 0.9 meV/atom and 0.7 meV/atom higher than the FCC, showing a much smaller difference than the DFT-PBE-D3 and DP-PBE-D3 results. 
Besides, the slope of the EOS at tensile stress area for Ag and Au deviates from the DFT-PBE-D3 and DP-PBE-D3 results. 
For Au, both DP-PBE-D3 and EAM-Zhou reproduces the relative stability of FCC and HCP structures successfully. 
The energy minima of HCP predicted by DFT-PBE-D3, DP-PBE-D3 and EAM-Zhou are 6.0 meV/atom, 3.4 meV/atom and 0.5 meV/atom higher than the FCC, respectively. 
The difference predicted by EAM-Zhou is one order of magnitude smaller than those predicted by DFT-PBE-D3 and DP-PBE-D3.  

\begin{center}
\begin{table}
\setlength{\abovecaptionskip}{0.cm}
\setlength{\belowcaptionskip}{0.9cm}
\caption{Surface formation energies for Ag and Au (100), (110), and (111) surfaces. }\label{tab:surf}
\resizebox{\textwidth}{!}{
\begin{threeparttable}
\begin{tabular}{c|cccccc|cccccc}
\hline
\hline
\multirow{2}{*}{Surface} & \multicolumn{6}{c}{Ag} & \multicolumn{6}{c}{Au} \\ 
\cline{2-13}																									
	 &	EAM-Zhou	&	PBE	&	DP	&	PBE-D3	&	DP-PBE-D3	&	Expt. 	&	EAM-Zhou	&	PBE	&	DP	&	PBE-D3	&	DP-PBE-D3	&	Expt.\\	
\hline																									
$E_{surf}$~(100)~($J/m^{2}$) 	&	0.97	&	0.84	&	0.84	&	1.39	&	1.23	&		&	1.08	&	0.87	&	0.86	&	1.56	&	1.46	&		\\
$E_{surf}$~(110)~($J/m^{2}$) 	&	1.11	&	0.91	&	0.87	&	1.49	&	1.31	&		&	1.24	&	0.87	&	0.87	&	1.68	&	1.56	&		\\
$E_{surf}$~(111)~($J/m^{2}$) 	&	0.91	&	0.73	&	0.71	&	1.33	&	1.13	&	1.25$^{a,b}$	&	0.95	&	0.71	&	0.67	&	1.41	&	1.25	&	1.5$^{a,b}$	\\

\hline
\end{tabular}
 \begin{tablenotes}
        \footnotesize
        \item[a] Reference\cite {Compton2014C}.
        \item[b] Reference\cite {de1988met}.
      \end{tablenotes}
\end{threeparttable}}
\end{table}
\end{center}

The unrelaxed surface formation energies $E_{surf}$ of pure Ag and Au are listed in Table~\ref{tab:surf}. 
It is defined as following: 
\begin{eqnarray}
E_{surf} = \frac{1}{2A} (E_{tot} - N_{s}E_{0}).
\label{eq:surf}
\end{eqnarray}
Here $E_{tot}$, $N_s$ and $A$ denote the energy of unrelaxed surface structure, the number of atoms and the surface area, respectively. 

The DFT-PBE-D3 surface formation energies are in good agreement with the experimental value, while the DFT-PBE predicts energies systematically lower than the experiment in both cases of Ag and Au. 
This indicates that the D3 dispersion correction is important in the quantitative prediction of the surface formation energies. 
The DP-PBE results agree well with the DFT-PBE predictions. 
The DP-PBE-D3 results are in reasonable agreement with the DFT-PBE-D3. 
More importantly, the DP-PBE-D3 predicts the correct relative stability of different surfaces. 
The EAM-Zhou potential describes the relative magnitudes reliably likewise.
However, the surface energies predicted by EAM-Zhou potential have a shift of about $-0.40$~J/m$^2$ for Ag and $-0.48$~J/m$^2$ for Au with respect to the DFT-PBE-D3 predictions.

\paragraph{Ag-Au alloys}

\begin{figure}
\begin{center}
\includegraphics[width=0.9\textwidth]{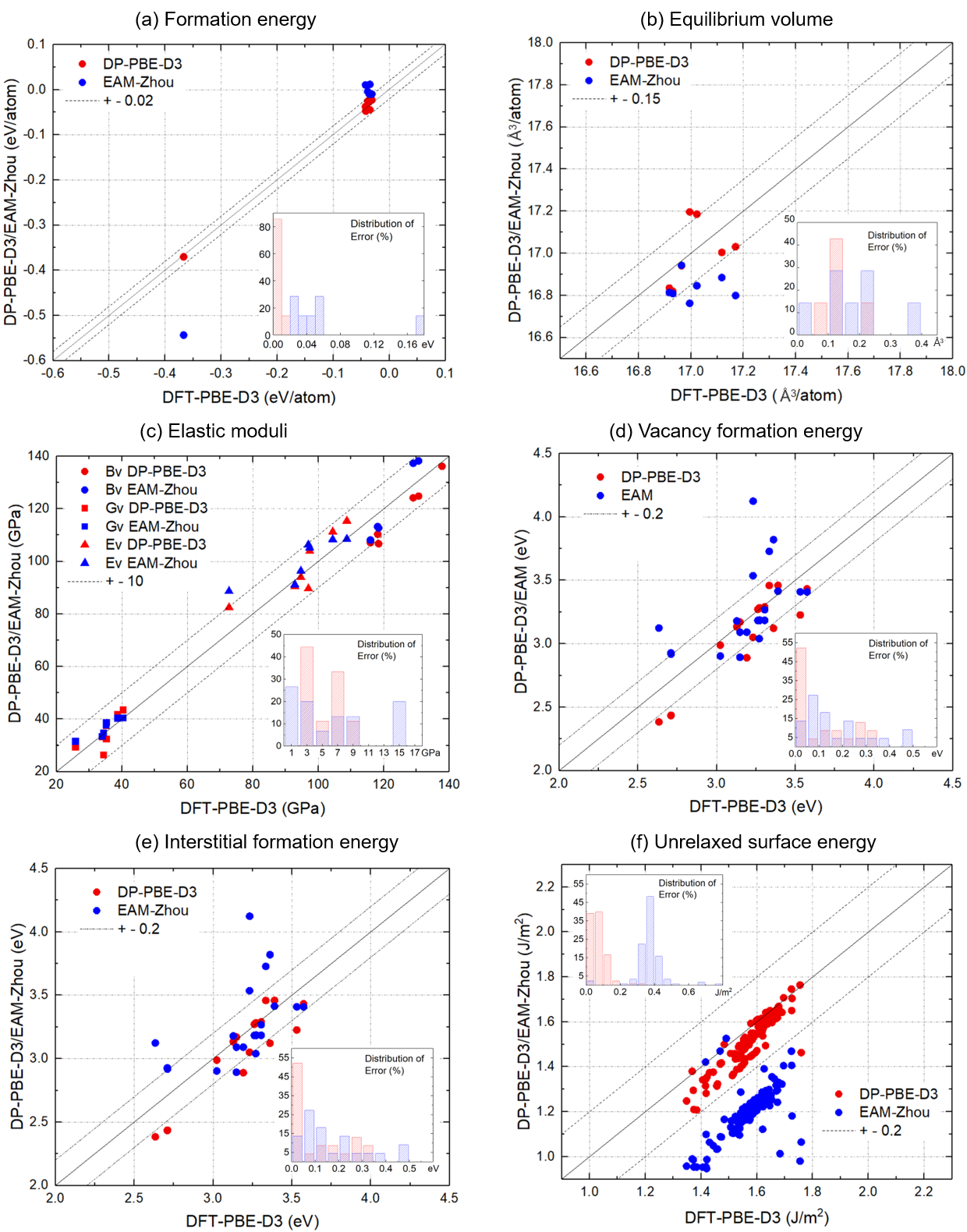}\\[4pt]  
\caption{Comparison of Ag-Au alloy properties predicted by DP-PBE-D3 and EAM-Zhou potentials, with respect to DFT-PBE-D3 results. 7 Structures are from the MP database. (a) 7 formation energies, (b) 7 equilibrium volumes per atom, (c) 21 elastic moduli, (d) 16 relaxed vacancy formation energies, (e) 23 relaxed interstitial formation energies and (f) 120 unrelaxed surface energies.
}\label{fig:AgAu}
\end{center}
\end{figure}

For the alloy systems, 7 crystalline Ag-Au alloy structures in the MP database\cite {jain2013commentary} are selected as test cases. 
None of these structures are explicitly included in the training data set. 
We compare results of DFT-PBE-D3, DP-PBE-D3 and EAM-Zhou potentials for the 7 alloy structures in Fig.~\ref{fig:AgAu} in the following items: (a) formation energy, (b) equilibrium volume, (c) elastic constants, (d) vacancy formation energies, (e) interstitial energies and (f) unrelaxed surface energies. 
The error distribution of the corresponding properties are included in each subplot.

The formation energy of the Ag-Au alloy is defined as
\begin{equation}
E^{formation}_{Ag-Au} = E_{Ag-Au}^{0}(x_{Ag})-x_{Ag}E^0_{Ag}-(1-x_{Ag})E^0_{Au},
\end{equation}
where $E^{0}_{Ag-Au}$ denotes the equilibrium energy per atom of the Ag-Au alloy structure. 
$x_{Ag}$ and $x_{Au}$ denote the component concentration of Ag and Au in the certain alloy structure, respectively.
$E^0_{Ag}$ and $E^0_{Au}$ are the equilibrium energies per atom of the pure metal in their equilibrium FCC states. 
The vacancy, interstitial and surface formation energies are calculated in the same way as the pure Ag and Au.
Some of the initial interstitial structures are nonphysical, e.g. the distance between two atoms is too close. 
These structures are difficult to converge in the VASP calculations, and are not considered in the following investigations.

In most of the test results, DP-PBE-D3 predictions agree well with the DFT-PBE-D3 reference results. 
For the formation energy, the absolute difference between the DP-PBE-D3 and DFT-PBE-D3 results is below 0.02 eV${\cdot}$atom$^{-1}$, as displayed in Fig.~\ref{fig:AgAu} (a).
Regarding the results of the EAM-Zhou potential, the largest deviation is 0.19 eV${\cdot}$atom$^{-1}$, obviously larger than the DP-PBE-D3 model. 
For the equilibrium volume displayed in Fig.~\ref{fig:AgAu} (b), the absolute errors of all the tested structures predicted by DP-PBE-D3 are smaller than 0.25~\AA$^3\cdot$atom$^{-1}$. 
By contrast, the equilibrium volume of MP-1183224 predicted by EAM-Zhou potential has an error of 0.38~\AA$^3$ $\cdot$ atom $^{-1}$, which is larger than the error of DP.
For the elastic moduli displayed in Fig.~\ref{fig:AgAu} (c), 
both DP-PBE-D3 and EAM-Zhou potentials are proper to describe the elastic properties, and they are of similar accuracy. 
Most of the absolute errors are beneath 10 GPa, except that the Young's modulus $E_v$ of the structure MP-985287 predicted by EAM-Zhou potential deviates from DFT-PBE-D3 by 16~GPa. 
For the vacancy and interstitial formation energies displayed in Fig.~\ref{fig:AgAu} (d-e), all DP-PBE-D3 predictions give absolute errors under 0.32 eV with respect to DFT-PBE-D3 values, while the largest absolute errors in EAM-Zhou simulations are 0.64 eV and 0.48 eV for vacancy and interstitial formation energies, respectively. 
The accuracy of DP is notably higher than EAM-Zhou in predicting the formation energy of point defects.
The results of the unrelaxed surface energies are displayed in Fig.~\ref{fig:AgAu} (f). 
The absolute errors from DP-PBE-D3 are beneath 0.2 $J/m^{2}$ with respect to the DFT-PBE-D3 values. 
In contrast, the EAM-Zhou-predicted surface formation energies present a systematical shift of $\sim -0.4$~J/m$^2$ compared with DFT-PBE-D3.

From the above analysis, we conclude that DP-PBE-D3 could predict the basic properties accurately, while the EAM-Zhou potential is less accurate in some cases. 
We emphasize that specific configurations, such as the vacancy, interstitial, or the alloy surface structures, are not included in the training data set explicitly.
The exploration step enlarged the sampled space of the DP-PBE-D3 model, which could predict the properties, to a larger extent, of special configurations, or more complicated defect structures.
The great performance of Ag-Au DP-PBE-D3 model stimulates our confidence to apply it in the investigations of the surface properties, surface segregation, and the surface adsorption and diffusion behaviors.

\subsection{\label{sec:recons}Au (111) surface reconstruction}

In this section, we present a comprehensive study of the Au (111) surface reconstruction phenomenon using the Ag-Au DP-PBE-D3 model. 
The reconstruction of the Au (111) surface is of particular research interest from both experimental and simulation perspectives. 
Remarkably, CO adsorption on Au (111) could promote the oxidation of alcohols \cite{rodriguez2012promoting}, which is counterintuitive since CO is usually considered as a poison in transition-metal electrocatalysis\cite{koper2009mechanisms}. 
Rodriguez et.~al.\cite{rodriguez2009unusual,rodriguez2010carbon,rodriguez2012promoting} argued that the enhancement electrocatalytic effect only exists on the Au (111) or (001) surfaces with hexagonal reconstruction.
Early experimental works on the Au (111) surface reconstruction, including the electron diffraction \cite{finch1935electron}, transmission electron microscopy (TEM) \cite{yagi1979surface}, transmission electron diffraction \cite{heyraud1980anomalous}, 
as well as scanning tunneling microscopy \cite{woll1989determination,barth1990scanning,tao1991observations,chen1998scanning}, suggested that the reconstruction resulted from the formation of a stacking-fault with ( 22 $\times$ $\sqrt{3}$ ) or ( 23 $\times$ $\sqrt{3}$ ) unit cells. 
Following from the work of LaShell et.~al.\cite{lashell1996spin}, a series of study by the angle resolved photoelectron spectroscopy indicated that the reconstruction of Au (111) surface results from the spin splitting, and the origin is spin-orbit coupling \cite{henk2003spin,reinert2003spin}. 
Existing experiments on all the noble metals have revealed that such phenomenon only exits in Au, but not in Cu or Ag \cite{nicolay2001spin,reinert2001direct}.

There have been several numerical studies based on either DFT or EFFs.
Recent DFT studies of the ( 22 $\times$ $\sqrt{3}$ ) reconstructed surfaces investigated the detailed atomic structures, and a work by Torres et.~al.\cite{torres2014density} focused on the relationship between the reconstruction energy and the slab size. 
They derived that the reconstruction is favored when $p$ > 10, and the surface reconstruction may alter the properties of Au, which shall be considered in modeling tasks. 
We notice that DFT simulations were restricted to a limited system size, which is partially alleviated by some MD simulations driven by EFFs at the price of sacrificing the accuracy.
For example, by constructing a one-dimensional double-sine-Gordon soliton model with known solutions, Ravelo et.~al.~calculated the phonon spectrum with remarkably agreement with the experiments \cite{ravelo1989molecular}, while the one-dimensional assumption may require a firm validation.
Ercolessi et.~al.~calculated ( 7 $\times$ $\sqrt{3}$ ), ( 11 $\times$ $\sqrt{3}$ ) and ( 23 $\times$ $\sqrt{3}$ ) reconstructed surface with a many-body model Hamiltonian, named ``glue'', and predicted that the ( 11 $\times$ $\sqrt{3}$ ) reconstructed surface was the most stable structure \cite{ercolessi1987surface}, which is contradictory to the experimental results.
Currently, no interatomic potentials could correctly predict the Au (111) reconstruction, despite the fact that the reconstruction is important and cannot be ignored in the systematic study of the surface properties.
 
In our simulation, the construction of the surface configurations is described below. 
The configurations contain 9 layers of the Au (111) plane, separated by vacuum of 20 \AA~in the [111] direction to avoid the spurious interaction. 
Periodic boundary conditions are applied in the other two directions: [1$\Bar{1}$0] and [11$\Bar{2}$]. 
For the unreconstructed surfaces, there are ( $p$ $\times$ 2$\sqrt{3}$ ) slabs, where $p$ represents the stacking times along the [1$\Bar{1}$0]. 
In this work, $p$ is increased from 5 to 37.
For the reconstructed surfaces, a ( 1 $\times$ 2$\sqrt{3}$ ) slab is added to the end of the surface, that ( $p$ $\times$ 2$\sqrt{3}$ ) reconstructed surface contains $p$ + 1 slabs. 
Then the geometry is fully relaxed with the DP-PBE-D3 model. 
NPT MD simulations are conducted at 573 K for 100 ps using a 1 fs time step. 
Then the system is cooled to 1K at a rate of 8.19 K/ps. 
Finally, the system is relaxed by the conjugate gradient energy minimization scheme.

\begin{figure}
\begin{center}
\includegraphics[width=0.6\textwidth]{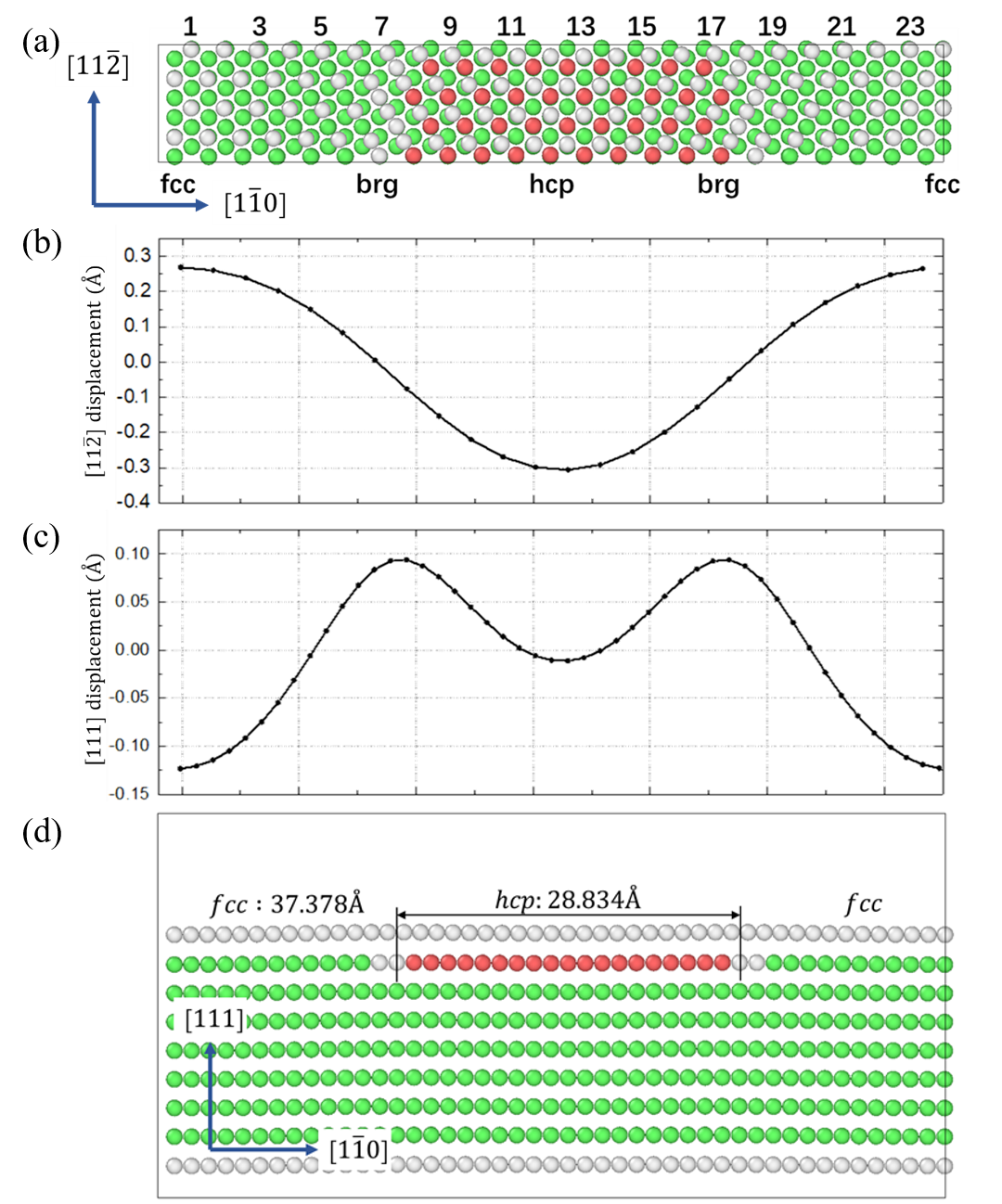}\\[3pt]  
\caption{(a) Schematic representation of the top view of the (23 $\times$ 2$\sqrt{3}$) reconstructed Au (111) surface. 
Atoms are colored in green and red if they are of FCC and HCP structure, respectively, and in white otherwise.
(b) Coordinations of the top layer in [11$\Bar{2}$] direction, and (c) in [111] direction. (d) Side view of the (23 $\times$ 2$\sqrt{3}$) reconstructed Au (111) surface. The widths of the FCC and HCP regions are assigned accordingly.
}\label{fig:Au111233}
\end{center}
\end{figure}

Fig.~\ref{fig:Au111233} illustrates the reconstructed $p$ = 23 surface. 
As illustrated by Fig.~\ref{fig:Au111233}~(a), the reconstructed surface shows a stacking-fault reconstruction: FCC region and HCP region with a bridge region connecting these two parts. 
The atoms in the topmost reconstructed layer displace not only along [1$\bar 1$0] but also notably along [11$\Bar{2}$] and [111] directions.
As displayed in Fig. ~\ref{fig:Au111233} (b), the reconstruction induces a lateral displacement, viz.~the HCP atoms are the lowest in [11$\Bar{2}$] direction and the FCC atoms are the highest. 
From Fig.~\ref{fig:Au111233} (c), the bridge region has the highest height in [111] direction, followed by HCP region, then the FCC region. 
Such phenomenon was observed in former experimental \cite{woll1989determination} and simulation results \cite {torres2014density}. 
The absolute length of the HCP and FCC regions are shown in Fig.~\ref{fig:Au111233} (d).
The FCC region is 37.38 \AA, while the HCP region is shorter with 28.83 \AA. 
This result agrees well with the former DFT calculations\cite{wang2007simulation,hanke2013structure}, and is consistent with the experimental observation that the FCC region is evidently larger than the HCP region\cite{barth1990scanning}.

To assess the stability of the reconstructed surfaces, the energy required to generate the reconstructed surface from the unreconstructed one, $E_{re}$, is evaluated as following:
\begin{eqnarray}
E_{re} = \frac{1}{2} (E_{r}-E_{u}-4E_{0})
\end{eqnarray}
$E_{r}$ and $E_{u}$ are the total energies of reconstructed and unreconstructed slabs, respectively. 
$E_{0}$ is the bulk energy of an Au atom.
Notice that this $E_{re}$ is for ( $p$ $\times$ $\sqrt{3}$ ). Negative $E_{re}$ represents that the reconstruction is favored at the corresponding $p$. 

\begin{figure}
\begin{center}
\includegraphics[width=0.98\textwidth]{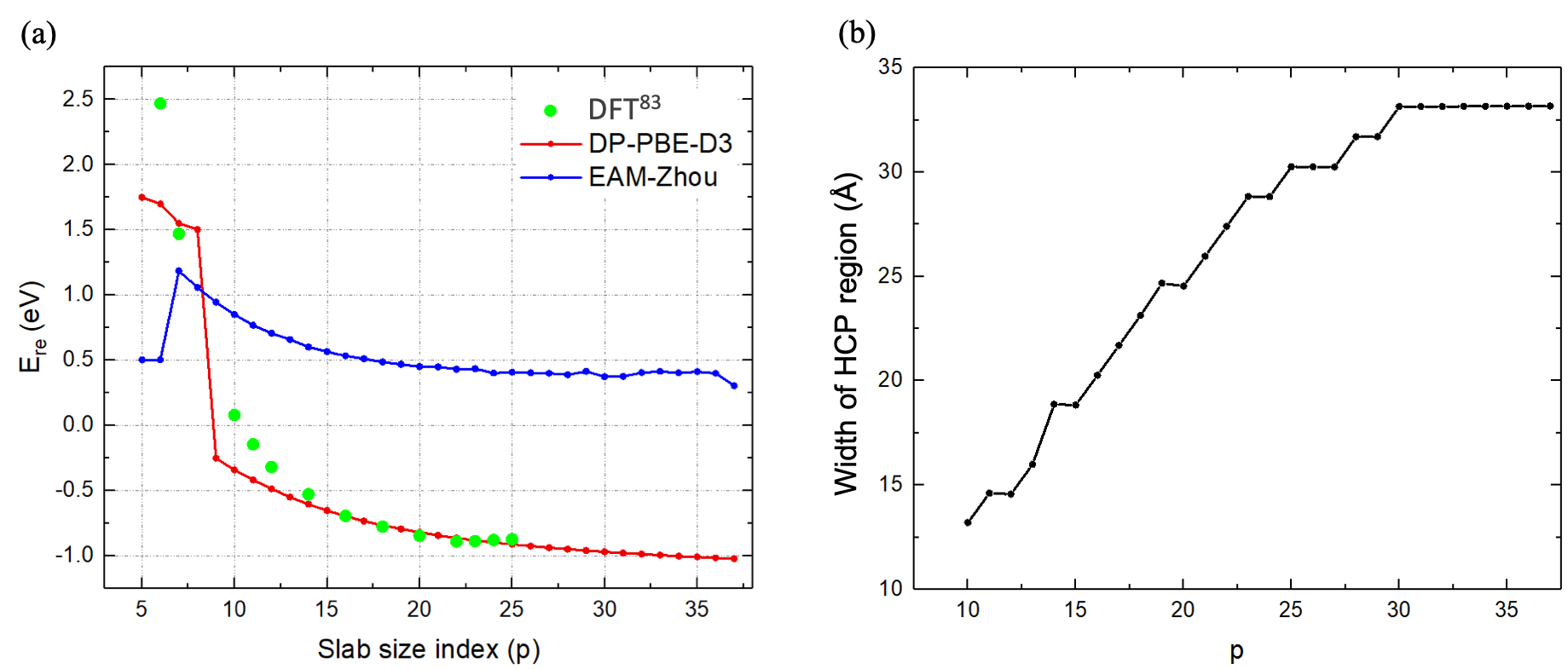}\\[4pt]  
\caption{(a) Reconstruction energy $E_{re}$ with increasing slab size index $p$. (b) The relationship between the widths of HCP region and $p$.
}\label{fig:Au111energy}
\end{center}
\end{figure}

Fig.~\ref{fig:Au111energy} (a) displays the reconstruction energy with different slab size index $p$ simulated by DP-PBE-D3 and EAM-Zhou. 
DFT results from Ref.\cite {torres2014density} are plotted in green dots for comparison.
The DP-PBE-D3 results fit quite well with the DFT results. 
When $p$ < 9, reconstruction is not favored due to a positive $E_{re}$ according to both DP-PBE-D3 and DFT results.
When $p$ > 10, both DP-PBE-D3 and DFT predicted $E_{re} < 0$, so reconstruction is preferred. 
The DP-PBE-D3 shows that as the slab size increases, the reconstruction energy decreases monotonically with a vanishing slope.
This can be understood by the fact that a larger reconstructed system has more degrees of freedoms to achieve a lower energy after the relaxation.
The DFT reconstruction energy firstly decreases, and then increases after $p=22$. 
The different trend should be attributed to the formidable computational expense of DFT that prevents the system from fully relaxed.
The EAM-Zhou potential predicts that the surface reconstruction is energetically unfavorable, which is qualitatively different from experimental observations and DFT calculations. 
In Fig.~\ref{fig:Au111energy} (b), the width of HCP regions as a function of the slab size $p$ is plotted. 
It is observed that  the width of the HCP region of the stable reconstructed surface converges to around 33.20 \AA\ at $p$ > 30. 
This is in a reasonable agreement with the experimentally observed HCP region width, $\sim 28$~\AA~\cite{barth1990scanning}.

In conclusion, 
since the Au (111) surface reconstruction is favored in large surfaces ($p$ > 10), the DP-PBE-D3 model, due to its  high accuracy and low computational cost, is a good candidate for the simulation of such phenomenon.
By contrast, DFT suffers from the difficulty of relaxing large reconstructed structures due to the high computational cost, while the EFF EAM-Zhou is qualitatively wrong in describing the stability of the reconstructed surfaces.

\subsection{\label{sec:seg} Segregation at the Ag-Au (111) surface}

In this section, we focus on simulating the segregation of the Ag on the Ag-Au (111) surfaces with different Ag concentration, using the Ag-Au DP-PBE-D3 model. 
The results of the segregation of Ag on the Ag-Au (111) surfaces predicted by DP-PBE-D3 model are compared with former DFT simulations~\cite{hoppe2017first}. 
Several experimental works focus on the influence of Ag towards the structures and the catalytic performance of the npAu. 
Wang et.~al.~\cite{wang2013catalytic} performed low-temperature CO oxidation experiments with npAu from Ag-Au and Cu-Au alloys and found that the activity of the npAu catalysts for CO oxidation scales linearly with the number of residual Ag surface atoms, which plays an important role in the catalytic performance. 
From TEM results, Fujita et.~al.~\cite{fujita2012atomic} found that the residual Ag in npAu stabilizes the surface defects, which are active sites for the catalytic oxidation of CO, through suppressing the (111) faceting kinetics. 
As a consequence, it is crucial to describe the segregation behaviors of Ag and Au in Ag-Au system correctly for further knowledge of the influence of the residual Ag in npAu on the catalytic activity. 
Considering the segregation of Ag towards the npAu surface, several experimental and theoretical investigations were carried out, yet varying results with great discrepancies were obtained. 
Attributing to the different thermodynamic conditions of the specimens and the calibration methods for Auger Electron Spectroscopy (AES) treatments, AES experiments on films~\cite{fain1974work,meinel1988segregation} or polycrystalline samples~\cite{overbury1976surface,bouwman1976surface} predicted different results, from Au enrichment on surface~\cite{fain1974work,bouwman1976surface} to Ag enrichment on surface~\cite{meinel1988segregation,overbury1976surface}. 
Low-energy Ion Scattering Spectroscopy studies~\cite{kelley1979surface} and Low Energy Electron Diffraction (LEED) analysis~\cite{derry2004comparison} resulted in the moderate Ag surface segregation. 
As for the theoretical simulations, conflicting results were reported with different methods. 
Bozzolo et. al. combined DFT and a semiempirical scheme called Bozzolo-Ferrante-Smith method, to study the bulk and surface properties of Ag-Au alloys~\cite{bozzolo2007atomistic}, while Deng et.~al.~worked on Ag-Au nanoparticles by a Modified Analytic Embedded Atom Method (MAEAM)~\cite{deng2011ag}. 
Both of them predicted the Ag enrichment in the surface. Contrarily, DFT studies predicted the Au enrichment in the surface~\cite{chen2008charge,paz2008structural,gould2014segregation}. 
Gould et.~al.~\cite{gould2014segregation} optimised the chemical orderings for cuboctahedral (AuAg)$_{147}$ nanoalloys by EFF and DFT, and found that the former preferred Ag enrichment in surface, while the latter preferred Au. 
Based on the detailed electronic analysis~\cite{bonavcic2002density,mitric2003structural}, the charges migrating towards the surface will be attracted by the Au atoms, since the Au atoms will draw electrons more easily when allocating on the surface. 
Hoppe et.~al., through a combined approach of DFT and statistical physics, explained the discrepancy between the DFT and experimental results by the residual pre-adsorbed oxygen atoms at the surface in the experiments, which lead to the Ag enrichment on the surface\cite{hoppe2017first}.
Overall, the empirical force field could not correctly predict the segregation of Ag or Au towards the surface, while the DFT method is limited in simulating the large scale of clusters. 
A reliable potential is required to describe the segregation of Ag or Au towards the surface, as well as the morphology and the structure of the npAu, and the related catalytic activities. 

\begin{figure}
\begin{center}
\includegraphics[width=0.3\textwidth]{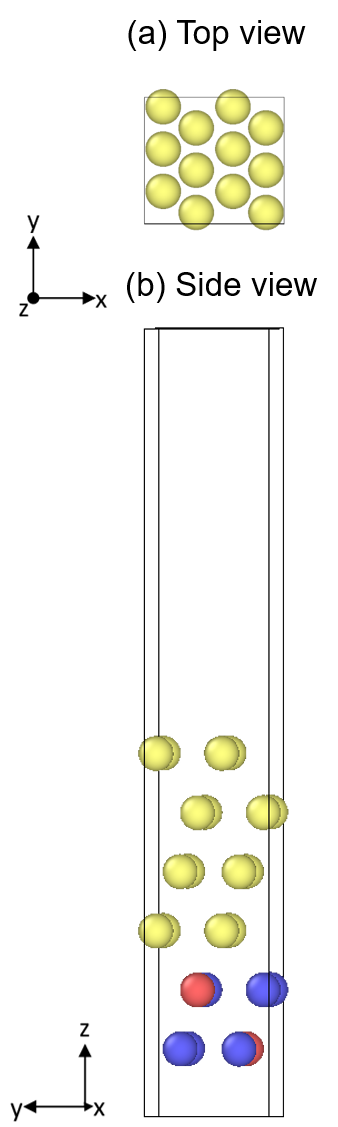}\\[1pt]  
\caption{Top (a) and side (b) views of the Ag-Au (111) surface layers with L1$_{2}$ ordered substrate. The Ag, Au atoms are colored in red and blue, respectively. The element types of atoms in the top four layers were allowed to change during the MC simulations. These atoms are colored in yellow to be distinguished with the L1$_{2}$ ordered substrate.  
}\label{fig:MC-mo}
\end{center}
\end{figure}

To investigate the segregation behavior of Ag-Au (111) surface, we applied the Monte Carlo (MC) simulations with the Metropolis algorithm~\cite{metropolis1953equation} implemented in \texttt{LAMMPS}. 
To get consistent results comparable with Ref.~\cite{hoppe2017first}, three different versions of (111) surface models are constructed.
Each of them contains 24 atoms in six layers: the two bottom layers are cut from pure Ag bulk, pure Au bulk and a L1$_{2}$ ordered bulk containing 25\% Ag, which is the ground state at very low temperature~\cite{ozolicnvs1998cu}. 
We define the two bottom layers as the substrate.
The composition of the substrate layers are fixed, while the atoms in the upper four layers are allowed to swap by the Metropolis rule during the MC simulation. 
Fig.~\ref{fig:MC-mo} represents the model with L1$_{2}$ ordered substrate employed in the MC simulations with (a) top view and (b) side view. 
The initial concentration of Ag in the upper four layers, $x^{tot}_{Ag}$, is set from 0.0625 to 0.9375 at a step of 0.0625.
To guarantee a random distribution of Ag and Au atoms, the initial temperature used in the Metropolis criterion is set to 100,000 K. 
An annealing schedule that is the same as that used in Ref.~\cite{hoppe2017first} is used: The temperature steps are set to $T$ = 100,000 K, 10,000 K, 2,000 K, 1,000 K, 10 K, and the decrements ($\delta$T) between the neighboring temperature steps are 10,000 K, 1,000 K, 200 K and 10 K, respectively.

\begin{figure}
\begin{center}
\includegraphics[width=0.8\textwidth]{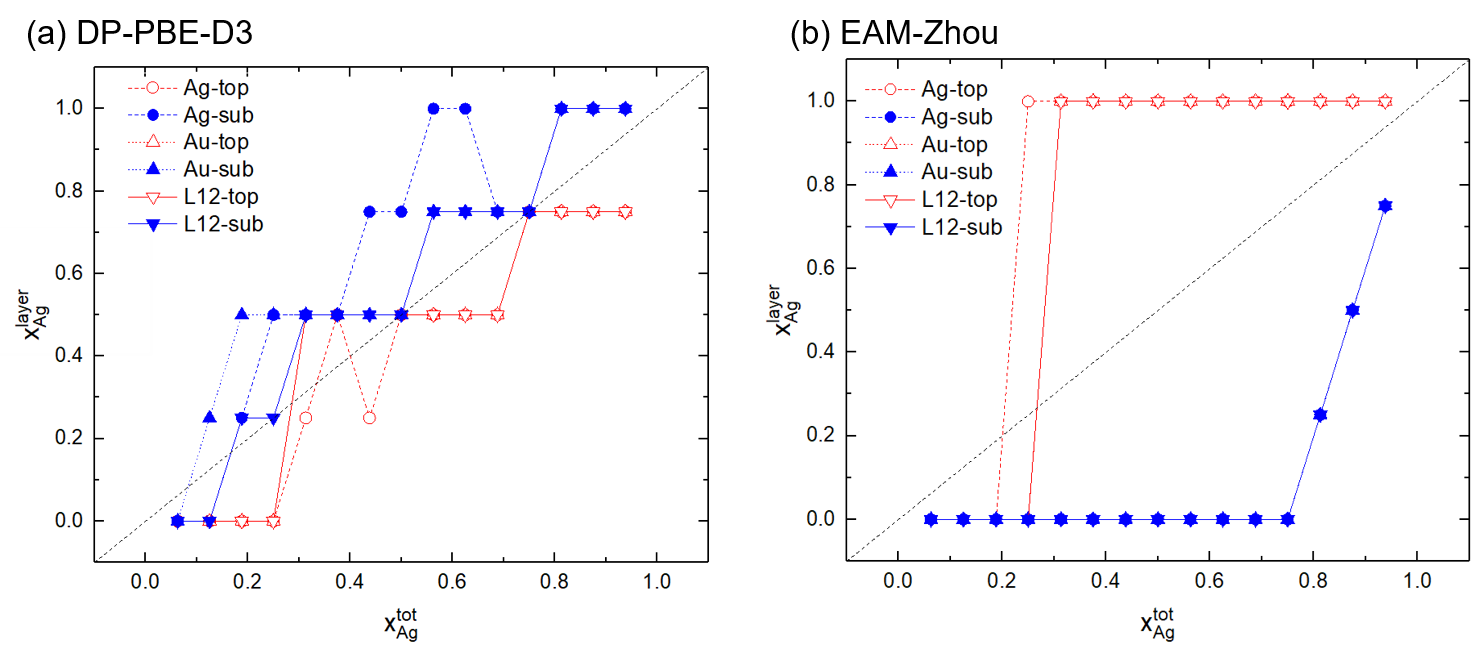}\\[2pt]  
\caption{Surface segregation profile for the Ag-Au (111) surface predicted by (a) DP-PBE-D3 and (b) EAM-Zhou with three different substrate structures: Ag, Au and L1$_{2}$ ordered bulk. The Ag concentrations in the top layer and sub-top layer, $x^{layer}_{Ag}$, are colored in red and blue, respectively. The substrate structures of Ag, Au and L1$_{2}$ ordered bulk are represented by circle, up-triangle and down-triangle. The black dashed line indicates no segregation, which is a guide to the eye.
}\label{fig:MC-re}
\end{center}
\end{figure}

Fig.~\ref{fig:MC-re} plots the relation between the concentration of Ag in the top layer (red lines) and sub-top layer (blue lines), denoted by $x^{layer}_{Ag}$, and the concentration of Ag in the upper four layers, denoted by $x^{tot}_{Ag}$, predicted by (a) DP-PBE-D3 and (b) EAM-Zhou.
The steps for the $x^{layer}_{Ag}$ is 0.25 due to the size of the cell. 
Generally, the DP-PBE-D3 predicts an Au enrichment in the top layer, which is consistent with DFT~\cite{hoppe2017first}, while the EAM-Zhou fails by predicting the Ag enrichment in the surface and the Au enrichment in the sub-top layer.
Although the tendency of segregation predicted by DP-PBE-D3 is satisfactory, some deviations from DFT require special attention.
Firstly,
we find that when one or two of the 16 atoms in the  upper four layers are Ag atoms, the stable positions of the Ag atoms are not in the top layer nor the sub-top layer.
In addition, for some $x^{tot}_{Ag}$ between 0.3 to 0.5, the DP-PBE-D3 predicts an enrichment of Ag in the top layer for the Au and L1$_{2}$ ordered substrates. 
We infer that this phenomenon rises from the limited size of the cell, which prevents the formation of Au rich top layer of a certain pattern.

Overall, the DP-PBE-D3 model predicts a comparable segregation behavior of Ag towards the Ag-Au surface as the DFT-PBE-D3 does, and is more reliable than the EAM-Zhou potential. Further, the DP-PBE-D3 model could be employed to investigate the structures of the npAu and characterize the Ag-Au (111) surfaces.

\subsection{\label{sec:dif} Adsorption and diffusion of adatom on Ag or Au surface}

Finally, we benchmark the performance of the DP-PBE-D3 model in describing the adsorption and diffusion of Ag(Au) adatom on Ag(Au) surface. 
The results are compared with a former DFT-PBE-D3 and EAM work\cite{bon2019reliability}. 
To get comparable results, the reconstruction of Au (111) surface is not considered. 

Experimental synthesis of the Ag-Au nanostructures always focuses on the surface control process and the complex evolution of the desired nanosystems \cite{qian2010cluster,jiang2011synergistic}. 
Meanwhile, theoretical treatment on Ag-Au alloy is often combined with the experiments for investigating the thermal and chemical properties \cite{ferrando2008nanoalloys}. 
For large cluster systems and time-consuming evolution procedures, MD \cite{baletto2003growth} or Monte Carlo simulations \cite{muller2005lattice} are required. 
To estimate the reliability of the potential in predicting the formation and growth of the nanoparticles, validations on the description of the adatom adsorption, diffusion, and further, the essential cluster structures, are required. 
Ahmad et.~al.~tested the reliabilty of two EAM models for describing the adsorption and diffusion of adatoms on the surfaces and clusters \cite{bon2019reliability}, and carried out the liquid-phase TEM to study the evolution of Ag-Au nanostructures, combining the MD simulations of the nanoalloy growth starting from Ag seeds\cite{ahmad2019template}. 
Although an overall agreement with the DFT results is observed, two available EAM potentials, EAM-Foils and EAM-Zhou \cite{foiles1986embedded,Zhou2004Misfit} show some deviations in the description of the diffusion and adsorption against DFT data. 
Therefore, a more accurate potential is required for more faithful descriptions and more detailed investigations.

\begin{figure}
\begin{center}
\includegraphics[width=0.5\textwidth]{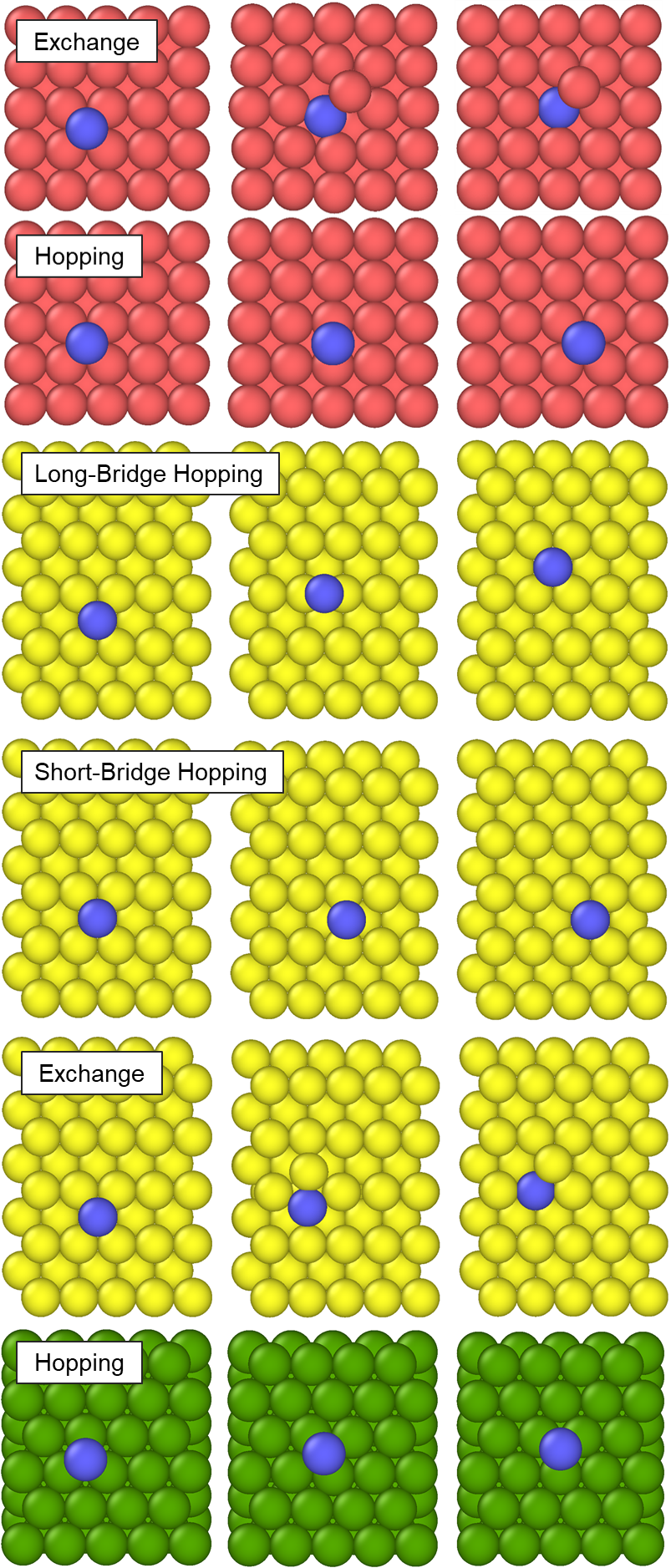}\\[2pt]  
\caption{Diffusion paths of the Ag (or Au) adatom (blue) on (100) (red), (110) (yellow) and (111) (green) Ag (or Au) surfaces. 
}\label{fig:diffusion}
\end{center}
\end{figure}

The surface structures of the Ag or Au systems are constructed as following. 
Considering the stacking sequence in the z direction, for the (100), (110) and (111) surfaces, 16, 22 and 18 layers were constructed, respectively. 
The size of the surface models in x and y directions for (100), (110) and (111) surfaces are 10 $\times$ 10, 10 $\times$ 10, and 14 $\times$ 8, respectively. 
Adsorption positions and diffusion paths are displayed in Fig.~\ref{fig:diffusion}. 
The Ag or Au atatoms located on hollow sites and the diffusion mechanisms are different for the surface indexes. 
On the (100) surface, we consider the Ag (Au) adatom diffusion by exchanging with the closest surface atom (Exchange), and by hopping to the closest equivalent site (Hopping). 
On the (110) surface, the exchange mechanism is also considered, while the hopping mechanism consists of long-Bridge hopping (LB hopping) and short-bridge hopping (SB hopping), with different distances for the adatom to diffuse. 
On the (111) surface, only the hopping mechanism is taken into consideration. 
The adsorption energies and diffusion barriers of the DFT-PBE-D3 and three potentials: EAM-Foil~\cite{foiles1986embedded}, EAM-Zhou and DP-PBE-D3 model are presented in Table.~\ref{tab:dif} . 

\vspace{5mm}
\begin{center}
\begin{table}[H]
\setlength{\abovecaptionskip}{0cm}
\setlength{\belowcaptionskip}{0.3cm}
\caption{Adsorption energies and diffusion barriers for Ag/Au adatoms on the Ag/Au (100), (110) and (111) surfaces. $E_{ads}$ is the adsorption energy, $E_{hop}$, $E_{exc}$, $E_{LB}$, $E_{SB}$ are the diffusion barriers of hopping, exchange, long-bridge and short-bridge mechanisms, respectively. The difference between DFT-PBE-D3 and EAM-Foils, EAM-Zhou and DP-PBE-D3 are listed in brackets, and the text of the results with the smallest deviation is bold. All units are in eV. }\label{tab:dif}
\resizebox{0.85\textwidth}{!}{
\begin{threeparttable}
\begin{tabular}{ccccccc}
\hline
\hline
Surface (100) & Adatom &  & DFT-PBE-D3$^{a}$ & EAM-Foils$^{a}$ & EAM-Zhou$^{a}$ & DP-PBE-D3 \\
\hline
\multirow{3}{*}{Ag} & \multirow{3}{*}{Ag} & $E_{ads}$ & 2.77 & 2.39 (0.38) & 2.19 (0.58) & \textbf{2.65 (0.12)} \\ 
\cline{3-7}
 & &	$E_{hop}$ & 0.40 & \textbf{0.48 (0.08)} & 0.50 (0.10) & 0.31 (0.09)	\\
\cline{3-7}
 &  &	$E_{exc}$ & 0.54 & 0.76 (0.22) & \textbf{0.66 (0.12)} & 0.69 (0.15)	\\
\hline
\multirow{3}{*}{Ag} & \multirow{3}{*}{Au} & $E_{ads}$ & 3.32 & \textbf{3.27 (0.05)} & 2.79 (0.53) & 3.57 (0.25) \\ 
\cline{3-7}
 &  &	$E_{hop}$ & 0.50 & 0.65 (0.15) & 0.71 (0.21) & \textbf{0.55	(0.05)} \\
\cline{3-7}
 &  &	$E_{exc}$ & 0.49 & 0.67 (0.18) & \textbf{0.47 (0.02)} & 0.62 (0.13)	\\
\hline
\multirow{3}{*}{Au} & \multirow{3}{*}{Au} & $E_{ads}$ & 3.60 & 3.49 (0.11) & 3.35 (0.25) & \textbf{3.52 (0.08)} \\ 
\cline{3-7}
 &  &	$E_{hop}$ & 0.59 & 0.67 (0.08) & 0.78 (0.19) & \textbf{0.61	(0.02)}\\
\cline{3-7}
 &  &	$E_{exc}$ & 0.09 & 0.35 (0.26) & 0.40 (0.31) & \textbf{0.13 (0.04)}	\\
\hline
\multirow{3}{*}{Au} & \multirow{3}{*}{Ag} & $E_{ads}$ & 3.12 & 2.70 (0.42) & 2.49 (0.63) & \textbf{2.80 (0.32)} \\ 
\cline{3-7}
 &  &	$E_{hop}$ & 0.47 & 0.57 (0.10) & 0.98 (0.51) & \textbf{0.38 (0.09)}	\\
\cline{3-7}
 &  &	$E_{exc}$ & 0.16 & 0.50 (0.34) & 0.60 (0.44) & \textbf{0.31 (0.15)}	\\
\hline
\hline
Surface (110) & Adatom & & DFT-PBE-D3$^{a}$ & EAM-Foils$^{a}$ & EAM-Zhou$^{a}$ & DP-PBE-D3 \\
\hline
\multirow{4}{*}{Ag} & \multirow{4}{*}{Ag} & $E_{ads}$ & 2.84 & 2.65 (0.19) & 2.56 (0.28) & \textbf{2.93 (0.09)} \\ 
\cline{3-7}
 &  &	$E_{LB}$ & 0.77 & \textbf{0.83 (0.06)} & 1.11 (0.34) & 0.87 (0.10)	\\
\cline{3-7}
 &  &	$E_{SB}$ & 0.30 & \textbf{0.32 (0.02)} & 0.25 (0.05) & 0.244 (0.056)	\\
\cline{3-7}
 &  &$E_{exc}$ & 0.26 & 0.42 (0.16) & 0.29 (0.03) & \textbf{0.239 (0.021)}	\\
\hline
\multirow{4}{*}{Ag} & \multirow{4}{*}{Au} & $E_{ads}$ & 3.53 & \textbf{3.59 (0.06)} & 3.25 (0.28) & 3.88 (0.35) \\ 
\cline{3-7}
 &  &	$E_{LB}$ & 0.83 & 1.13 (0.30) & 1.44 (0.61) & \textbf{0.98 (0.15)}	\\
\cline{3-7}
 &  &	$E_{SB}$ & 0.38 & \textbf{0.30 (0.08)} & \textbf{0.30 (0.08)} & 0.188 (0.192)	\\
\cline{3-7}
 &  &	$E_{exc}$ & 0.28 & 0.36 (0.08) & \textbf{0.25 (0.03)} & 0.186 (0.094)	\\
\hline
\multirow{4}{*}{Au} & \multirow{4}{*}{Au} & $E_{ads}$ & 2.59 & 3.71 (1.12) & 3.65 (1.06) & \textbf{3.63 (1.04)} \\ 
\cline{3-7}
 &  &	$E_{LB}$ & 0.72 & 1.04 (0.32) & 1.26 (0.54) & \textbf{0.88 (0.16)}	\\
\cline{3-7}
 &  &	$E_{SB}$ & 0.36 & 0.25 (0.11) & \textbf{0.29 (0.07)} & 0.44 (0.08)	\\
\cline{3-7}
 &  & $E_{exc}$ & 0.38 & 0.40 (0.02) & \textbf{0.38 (0.00)} & 0.28 (0.10)	\\
\hline
\multirow{4}{*}{Au} & \multirow{4}{*}{Ag} & $E_{ads}$ & 2.99 & \textbf{2.91 (0.08)} & 2.78 (0.21) & \textbf{3.07 (0.08)} \\ 
\cline{3-7}
 &  &	$E_{LB}$ & 0.76 & 1.13 (0.37) & 1.45 (0.69) & \textbf{0.80 (0.04)}	\\
\cline{3-7}
 &  &	$E_{SB}$ & 0.33 & 0.30 (0.03) & 0.30 (0.03) & \textbf{0.32 (0.02)}	\\
\cline{3-7}
 &  &	$E_{exc}$ & 0.31 & 0.70 (0.39) & 0.24 (0.07) & \textbf{0.27 (0.04)}	\\
\hline
\end{tabular}
\end{threeparttable}}
\end{table}
\end{center}

\begin{center}
\begin{table}[H]
\resizebox{0.85\textwidth}{!}{
\begin{threeparttable}
\begin{tabular}{ccccccc}
\hline
\hline
Surface (111) & Adatom & & DFT-PBE-D3$^{a}$ & EAM-Foils$^{a}$ & EAM-Zhou$^{a}$ & DP-PBE-D3 \\
\hline
\multirow{2}{*}{Ag} & \multirow{2}{*}{Ag} & $E_{ads}$ & 2.47 & 2.14 (0.33) & 1.91 (0.56) & \textbf{2.43 (0.04)} \\ 
\cline{3-7}
 &  &	$E_{hop}$ & 0.06 & \textbf{0.06 (0.00)} & 0.05 (0.01) & \textbf{0.06 (0.00)}	\\
\hline
\multirow{2}{*}{Ag} & \multirow{2}{*}{Au} & $E_{ads}$ & 3.12 & 2.92 (0.20) & 2.37 (0.75) & \textbf{3.26 (0.14)} \\ 
\cline{3-7}
 &  &	$E_{hop}$ & 0.09 & 0.05 (0.04) & 0.04 (0.05) & \textbf{0.07 (0.02)}	\\
\hline
\multirow{2}{*}{Au} & \multirow{2}{*}{Au} & $E_{ads}$ & 3.19 & \textbf{3.03 (0.16)} & 2.86 (0.33) & 3.00 (0.19) \\ 
\cline{3-7}
 &  &	$E_{hop}$ & 0.12 & 0.02 (0.10) & 0.05 (0.07) & \textbf{0.09 (0.03)}	\\
\hline
\multirow{2}{*}{Au} & \multirow{2}{*}{Ag} & $E_{ads}$ & 2.51 & 2.35 (0.16) & 2.20 (0.31) & \textbf{2.47 (0.04)} \\ 
\cline{3-7}
 &  & $E_{hop}$ & 0.09 & 0.05 (0.04) & 0.05 (0.04) & \textbf{0.09 (0.00)}	\\
\hline
\end{tabular}
 \begin{tablenotes}
        \footnotesize
        \item[a] Reference\cite {bon2019reliability}.
      \end{tablenotes}
\end{threeparttable}}
\end{table}
\end{center}

The adsorption energy $E_{ads}$ is defined by:
\begin{eqnarray}
E_{ads} = E_{total}-E_{surf}-E_{atom},
\end{eqnarray}
where $E_{total}$, $E_{surf}$ and $E_{atom}$ represents the energy of the surfaces with and without the adatom, and the isolated adatom, respectively. 
The diffusion barriers are calculated by the nudged elastic band (NEB) method \cite{2000A,2000Improved}, which is implemented in \texttt{LAMMPS}. 
Eight replicas are employed in each process and the spring constant is set as 1.0 eV/\AA$^{2}$. 
The stopping tolerances of structure relaxation are $10^{-10}$ eV and $10^{-15}$ eV/\AA\ for energy and force, respectively.
For convenience, we use (A/B) to denote the situation that a B adatom is located on the A surface, e.g. (Ag/Au) means an Au adatom on Ag surface. 

For the (100) surface, the $E_{ads}$ predicted by DP-PBE-D3 is the most accurate one, except the (Ag/Au) case with an error of $\sim7.53\%$. 
Considering the diffusing mechanism, DFT-PBE-D3 predicts that the hopping mechanism is preferred in the (Ag/Ag) case, while the exchanging mechanism is preferred in the other three cases. 
DP-PBE-D3 predicts the right tendency except for the (Ag/Au) case, for the diffusion barriers of hopping and exchanging mechanisms predicted by DFT-PBE-D3 in the (Ag/Au) case is quite close.
EAM-Foils also predicts the right tendency except the (Ag/Au) case, while EAM-Zhou predicts the right tendency for all four cases. 
Although the preferred mechanism predicted by the EAM-Foil, EAM-Zhou and DP-PBE-D3 are almost correct, the accuracy of the barriers worth discussions. 
We calculate the root-mean-square error (RMSE) of the diffusion barriers of all the 8 cases (i.e.~the hoping and exchanging barriers of Ag/Ag, Ag/Ag, Au/Au and Au/Ag).
The EAM-Foil, EAM-Zhou and DP-PBE-D3 give RMSEs of 0.197~eV, 0.286~eV, and 0.110~eV, respectively. Obviously the accuracy of the DP-PBE-D3 is better than the EAM potentials.

For the (110) surface, the $E_{ads}$ predicted by DP-PBE-D3 is the most accurate one comparing with EAM-Foil and EAM-Zhou, except the (Ag/Au) case with an error of $\sim9.92\%$. 
Considering the diffusing mechanism, DFT-PBE-D3 predicts that the short-bridge hopping mechanism is preferred in (Au/Au) case, while the exchanging mechanism is preferred in other three cases.
DP-PBE-D3 predicts the exchanging mechanism is preferred for all four cases, so only the (Au/Au) case is wrong. 
EAM-Foils predicts the short-bridge hopping mechanism as the preferred one for all four cases, yet only the (Au/Au) case agrees with the DFT-PBE-D3. 
EAM-Zhou predicts the right tendency except the (Ag/Ag) case.
We also calculate the RMSE of the diffusion barriers in the 12 cases to compare the accuracy of the three potentials, and the RMSE is 0.259~eV, 0.399~eV, and 0.126~eV for EAM-Foil, EAM-Zhou and DP-PBE-D3, respectively. The accuracy of the DP-PBE-D3 is better than EAM-Foil and EAM-Zhou.

For the (111) surface, the $E_{ads}$ predicted by DP-PBE-D3 is the most accurate one comparing with EAM-Foil and EAM-Zhou, except the (Au/Au) case with an error of $\sim5.96\%$. The barriers of the hopping mechanism predicted by DFT-PBE-D3 are all more accurate than the EAM-Foils and the EAM-Zhou in all four cases. Noticing that the diffusion barriers on the (111) surface are evidently smaller than on the (100) and (110) surface, the DP-PBE-D3 model distinguished the tiny difference rarely, which is better than the EAM-Foils and EAM-Zhou potentials.

To summarize, the performance of DP-PBE-D3 is better than the two EAM potentials evidently. Our DP-PBE-D3 model is reliable in predicting the adsorption and diffusion of Ag/Au atoms on the Ag/Au (100), (110) and (111) surfaces. 

\section{Conclusion}\label{sec:conclusion}

In this work, to model the Ag-Au alloys in the full concentration space, a DP-PBE-D3 model is generated using the concurrent learning scheme DP-GEN in combination with a transfer learning process.
The DP-PBE-D3 model is benchmarked with EOS, elastic moduli, point defect formation energy and the surface formation energy, and shown to be of comparable accuracy with the DFT-PBE-D3 (DFT calculation with PBE exchange correlation functional and D3 dispersion correction). 
We further demonstrate the reliability of the DP-PBE-D3 model in three typical surface questions, the Au (111) surface reconstruction, the segregation at Ag and Au atoms at the (111) surface of Ag-Au alloys and the adsorption and diffusion of adatoms on the Ag or Au surfaces. 
The DP-PBE-D3 model reproduces the experimental observations with a semi-quantitative accuracy, which is superior to the state-of-the-art EAM potentials that are qualitatively wrong in the first two cases and are of sub-optimal accuracy in the last case.
Since the training dataset does not explicitly include any configuration from the applications, the DP-PBE-D3 model is thus transferable in the demonstrative applications, and could be employed in further investigation of the Ag-Au surface diffusion properties, the structures of npAu and the generation and growth processes of the nano-particles and cluster. 

\section*{Acknowledgment}
The work of Y.N. Wang is supported by the fellowship of China Postdoctoral Science Foundation under Grant No.~2020M680733.
The work of H.W. is supported by the National Science Foundation of China under Grant No.~11871110.
The work of H.W. and L.Z. is supported in part by Beijing Academy of Artificial Intelligence (BAAI).

\newpage

\end{document}